# The Insulin-RB Synapse in Health and Disease: Cellular Rocket Science

**Razvan Tudor Radulescu**

Molecular Concepts Research (MCR), Munich, Germany

E-mail: ratura@gmx.net

Motto:   "It is far better even to foresee without certainty than not to foresee at all." (Henri Poincaré)

Keywords:   physiology, pathology, cell growth regulation, insulin, retinoblastoma protein (RB), insulin-degrading enzyme (IDE), tumor suppressor, embryogenesis, wound healing, insulin resistance, diabetes mellitus, obesity, hyperinsulinemia, neurodegeneration, Alzheimer´s disease, pain modulation, analgesia, cancer, oncogenesis, senescence, aging, treatment, therapy, MCR peptide, nucleocrine, molecular synapse, cybernetics, information, force, velocity, speed, rocket science

# ABSTRACT

Time has come for a survey of our knowledge on the physical interaction between the growth-promoting insulin molecule and retinoblastoma tumor suppressor protein (RB). Theoretical and experimental observations over the past 15 years reviewed here indicate that the insulin-RB dimer may represent an essential molecular crossroads involved in major physiological and pathological conditions. Within this system, the putative tumor suppressor insulin-degrading enzyme (IDE) should be an important modulator. Perhaps most remarkably, the abstraction of this encounter between insulin and RB, two growth-regulatory giants acting either in concert or against each other depending on the respective cellular requirements, reveals that Nature may compute in controlling cell fate and we could follow in its footsteps towards developing more efficient therapeutics as well as novel technical devices.

**Insulin meets RB: synopsis of a discovery**

Biological information is encoded in individual genes and proteins, yet life emerges from multiple *interactions* among various such molecules. A prominent paradigm for these events is the docking of ligands floating in the extracellular space to receptors located on the surface of cells. Within this category, one example of central importance to cell metabolism is the complex formation between insulin and the insulin receptor.

Investigations carried out over the past 15 years, however, revealed that, besides binding and thereby recruiting the (cell membrane) insulin receptor as a molecular switchboard for signal transduction, insulin likely targets a second crucial molecule in order to decisively influence cell fate: the (nuclear) retinoblastoma tumor suppressor protein (RB). Accordingly, it was initially shown that insulin harbors the LXCXE RB-binding motif (1) and subsequently demonstrated that insulin indeed physically interacts with RB, specifically the latter´s predicted insulin binding site synthesized as a peptide (2), and also the full-length RB protein in the context of living human tumor cells (3-5). Moreover and interestingly, the predominantly nuclear contact formation between insulin and RB was shown to correlate with an increase in cell proliferation (3), suggesting that, similar to RB-binding viral oncoproteins, the attachment of insulin to RB inhibits RB´s antiproliferative function.

The potentially paramount importance of the insulin-RB encounter is further underscored by a third player likely to impact on this growth-regulatory system: insulin-degrading enzyme (IDE). In 1994, it was initially hypothesized that IDE may exert tumor-suppressive activity by preventing that insulin moves into the nucleus and subsequently neutralizes RB, the net effect being a maintenance of natural cancer prophylaxis by functional RB (6).

A decade later, a striking homology between IDE and RB was identified which indicated an unexpected capacity of RB to act as an insulin-degrading protease by virtue of a zinc-binding catalytic site similar to that present in IDE and located in the carboxyterminal moiety of RB (7). Conversely, this structural kinship conferred further plausibility to the above 1994 hypothesis on IDE being a putative tumor suppressor.

Intriguingly, a first experimental study on IDE in human cancer tissues appears to confirm the validity of this concept as IDE was shown to be overexpressed in breast cancer vs. normal breast, yet underexpressed in lymph node metastases vs. primary tumors (8), consistent with the notion of a gradual loss of IDE tumor-suppressive activity with progressive oncogenesis and similar to the biphasic expression patterns of Bax (9) and HtrA1 (10).

It is noteworthy that this first published demonstration of IDE in human malignancies had been preceded by extensive investigations gradually eliminating potential staining pitfalls (11) and thus establishing the reliability of the finally employed immunohistochemical method (8).

While the above theoretical and experimental studies on the potentially antineoplastic role of IDE opened the door a crack into this direction, a most recent insight pushed it wide open: it was the recognition that *insulin antagonism* may be a more widespread phenomenon as a natural means to protect (nuclear) RB and shared by proteins mechanistically as diverse as IDE, phosphatase and tensin homologue (PTEN) and insulin-like growth factor binding protein 7 (IGFBP-7), hence providing a heretofore unknown **second principle** besides cyclin-dependent kinase inhibition through p16, p21 and/or p27 for maintaining the antiproliferative activity of RB (12).

As if such a variety of insulin-antagonistic RB protectors had not sufficed, Nature seems to have endowed also RB itself with the propensity for self-defense. Beyond the above mentioned potential for RB to degrade insulin (7), RB may leave the cell by virtue of a putative N-terminal signal peptide (13) and interfere with the docking of insulin to its cell membrane receptor in the extracellular compartment (14), thereby preventing the initiation of phosphorylation cascades which would otherwise inactivate nuclear RB.

It is, however, also conceivable that this proposed RB interference with insulin signal transduction may not require RB secretion since intracellular RB could also avert intracrine insulin receptor activation by internalized or *de novo* synthesized insulin, thus abrogating a growth-stimulatory loop similar to short circuits previously described for isoforms of PDGF (15) and FGF (16) as well as for IL-3 (17).

**The insulin-RB heterodimer and its modulators in specific conditions**

Since there is no major protein-protein interaction relevant only in pathology, one should first consider the potential role of the insulin-RB complex in physiology. As such, it could be surmised that an increased quantity of insulin-RB heterodimers contributes not only to the rapid phases of embryonic and fetal development, but also to tissue regeneration following injury. The latter may be particularly applicable within the liver which is known to be an organ with a high capacity for self-renewal and under circumstances where insulin is employed therapeutically to accelerate wound closure and thereby restore tissue integrity (18). Consequently, a potentially oncogenic molecule such as insulin would be capable of cooperating with RB- by recruiting its cell survival-promoting function (19) while interfering with its antiproliferative actions- towards a physiological goal, thus resembling the Ras-RB cooperation during cell differentiation (20).

By contrast, under other circumstances, a predominance of this dimer may be deleterious. Accordingly, the natural tropism of insulin towards RB, if unbalanced, could contribute to the insulin resistance syndrome including type 2 diabetes mellitus and obesity as well as ultimately accelerate the aging process (21). Moreover, excessive insulin-RB encounters are likely to represent a major signal for carcinogenesis (1,3,22). Briefly, overflowing insulin predictably keeps RB from performing its anti-cancer and anti-aging tasks (14, 21, 23).

Beyond aging and cancer, the insulin-RB complex may be involved in the pathogenesis of various neurodegenerative conditions such as Alzheimer´s disease (AD). Foregoing this possible involvement would be a state of hyperinsulinemia- an aberrancy known to be associated with AD- causing a reactive increase in the number of insulin receptors and, upon their activation, an upregulation of IDE (24) that, with further disease progression, largely fails to fulfill its task to degrade insulin (and amyloid). As a result, internalized insulin could undergo unhindered translocation to the nucleus with subsequent binding of a significant amount of RB molecules. Given the post-mitotic nature of neurons, a predominance of insulin-RB nuclear contacts should then lead to cell death, not proliferation.

Most recently, it became apparent that the insulin-RB complex may also enhance pain signalling (25). The basis for this assumption is the fact that an analgesic spider venom peptide termed psalmotoxin 1 (26) was found to contain the PKT amino acid sequence in duplicate (27), i.e. a tripeptide identical to the signature postulated to enable RB to interfere with insulin´s binding to its receptor (14), suggesting that RB could block not only insulin´s pro-aging effects (14), but also its likely potentiation of pain signal transmission through PI3kinase (28).

Similar to the insulin-RB aging model in which insulin retains and inactivates RB in the nucleus such that the latter can no longer interfere with insulin-insulin receptor association, thus promoting aging (21), excessive insulin-RB dimerization would also abrogate RB´s contribution to analgesia and result in hyperalgesia.

Therefore, compounds known to antagonize the physical interaction between insulin and RB such as MCR peptides should be able to block not only cancer growth (3,5,29-35), but equally counteract organismal senescence (21), attenuate neurodegenerative disorders (36) and reduce pain sensation (37).

**General implications and outlook**

The question is then why Nature selected for the encounter between insulin and RB in various major conditions, physiological or pathological, or, more generally speaking, for nucleocrine communication, the insulin-RB dimer representing its paradigm (38), yet the EGF precursor-p130 complex being also an important example (39). Interestingly, a novel view may provide both the likely answer and open up a largely uncharted territory for future investigations. Accordingly, the insulin-RB complex may be best understood in terms of a novel theory for (biological) information transfer governed by the following equation:

$$F_i = m\, a_i$$

where $F_i$ is the cybernetic force affecting a given (biological) system (e.g. a cell), $m$ is the mass of a given cue (e.g. a biologically active molecule) and $a_i$ is its cybernetic acceleration.

The cybernetic acceleration $a_i$ itself may be defined as $a_i = 1/n$ where $n$ is the number of intermediary objects required for a cue to change the state of a system (e.g. for translating the message of a biological molecule into a modification of gene expression). Specificity of such accelerated information transfer would then reside in the structure of the cue, e.g. in the amino acid sequence of a signalling protein or peptide.

In keeping with this definition, for those insulin molecules present in insulin-RB dimers, $n$ is to be assigned a value of 1 since only RB, i.e. one molecule, mediates between insulin and DNA which in turn would yield the maximum $a_i$ value of 1 (more precisely, $a_i = 1/1 = 1$) and, ultimately, in conjunction with insulin´s $m$ value, the maximally achievable cybernetic force $F_i$. In other words, whenever there is no intermediary or yet only one intermediary to the gene-regulatory action of a molecule, this embodies maximal cybernetic simplicity and thus efficiency.

By contrast, if an extracellular ligand docking to a cell membrane receptor needs 5 intermediary molecules for its information to be relayed to the genes in the cell nucleus such that ultimately gene expression is switched on or off, this would yield a value of 1/5 or 0.2 for $a_i$ of the given ligand.

In this context, a special case is, for instance, insulin since this hormone and growth factor may both activate a cell membrane receptor and hence entrain a post-receptor signal transduction cascade involving many molecules, yet also translocate to the nucleus to bind RB. In this case, one would have to distinguish the two functional isoforms and assign to them distinct $F_i$ values accordingly.

In the light of these new considerations, the insulin-RB complex translates into maximal velocity for the transmission of insulin-enciphered biological information, and, as such, is a first candidate for driving cell proliferation within the cybernetics of cancer (40). Notably, my present perception is consistent with the concept of "regulated protein translocation" as a source of specificity and acceleration in biological signal transduction (41).

As knowledge penetrates into increasingly microscopic dimensions, I also propose to perceive the key contact formation occurring between insulin and RB as a (molecular) synapse, hence paralleling the relatively more macroscopic synapses between neurons or yet those created within the immune system by dendritic cells and T lymphocytes.

Due to its implications for biology and beyond, further thinking into the insulin-RB complex and congeners could push the cybernetic frontier far ahead. Ultimately, a new discipline may arise, akin to rocket science and thus contrasting the "black box" of conventional transcriptional activators (42).

**Acknowledgements**

# References

1. **Radulescu, R.T., and C.M. Wendtner.** 1992. Proposed interaction between insulin and retinoblastoma protein. *J. Mol. Recognit.* **5:** 133-137.
2. **Radulescu, R.T., M.R. Bellitti, M. Ruvo, G. Cassani, and G. Fassina.** 1995. Binding of the LXCXE insulin motif to a hexapeptide derived from retinoblastoma protein. *Biochem. Biophys. Res. Commun.* **206:** 97-102.
3. **Radulescu, R.T., E. Doklea, K. Kehe, and H. Mückter.** 2000. Nuclear colocalization and complex formation of insulin with retinoblastoma protein in HepG2 human hepatoma cells. *J. Endocrinol.* **166:** R1-R4.
4. **Radulescu, R.T., and J. Schulze.** 2002. Insulin-retinoblastoma protein (RB) complex further revealed: intracellular RB is recognized by agarose-coupled insulin and co-immunoprecipitated by an anti-insulin antibody. *Logical Biol.* **2:** 2-7.
5. **Radulescu, R.T., and K. Kehe.** 2007. Antiproliferative MCR peptides block physical interaction of insulin with retinoblastoma protein (RB) in human lung cancer cells. *arXiv*:0706.1991v1 [q-bio.SC]
6. **Radulescu, R.T.**, unpublished observation.
7. **Radulescu, R.T.** 2005. Zinc-binding motif similarity between retinoblastoma protein (RB) and insulin-degrading enzyme (IDE): insulin degradation as a potential tumor suppression principle. *Logical Biol.* **5:** 3-6.
8. **Radulescu, R.T., C. Hufnagel, P. Luppa, H. Hellebrand, W.-L. Kuo, M.R. Rosner *et al.*** 2007. Immunohistochemical demonstration of the zinc metalloprotease insulin-degrading enzyme in normal and malignant human breast: correlation with tissue insulin levels. *Internat. J. Oncol.* **30:** 73-80.
9. **Jansson, A., and X.F. Sun.** 2002. Bax expression decreases significantly from primary tumor to metastasis in colorectal cancer. *J. Clin. Oncol.* **20:** 811-816.
10. **Baldi, A., A. De Luca, M. Morini, T. Battista, A. Felsani, F. Baldi, *et al.*** 2002. The HtrA1 serine protease is down-regulated during human melanoma progression and represses growth of metastatic melanoma cells. *Oncogene* **21:** 6684-6688.
11. **Radulescu, R.T., A. Jahn, D. Hellmann, and G. Weirich.** 2007. Immunohistochemical pitfalls in the demonstration of insulin-degrading enzyme in normal and neoplastic human tissues. *arXiv*:0705.0374v1 [q-bio.TO]
12. **Radulescu, R.T.** 2007. One for all and all for one: RB defends the cell while IDE, PTEN and IGFBP-7 antagonize insulin and IGFs to protect RB. *Med. Hypotheses* **69:** 1018-1020.
13. **Radulescu, R.T.** 2004. Signal peptide-like sequence in retinoblastoma protein (RB): the signature for the secretion of a nuclear tumor suppressor. *Logical Biol.* **4:** 81-83.
14. **Radulescu, R.T.** 2003. Potential of retinoblastoma protein to block insulin receptor activation by insulin: structural and experimental clues to a novel anti-dogma on a dual inhibition of cancer and ageing. *Logical Biol.* **3:** 40-42.
15. **Bejcek, B.E., D.Y. Li, and T.F. Deuel.** 1989. Transformation by c-sis occurs by an internal autoactivation mechanism. *Science* **245:** 1496-1499.
16. **Acland, P., M. Dixon, G. Peters, and C. Dickson.** 1990. Subcellular fate of the int-2 oncoprotein is determined by choice of the initiation codon. *Nature* **343:** 662-665.

17. **Dunbar, C.E., T.M. Browder, J.S. Abrams, and A.W. Nienhuis.** 1989. COOH-terminal-modified interleukin-3 is retained intracellularly and stimulates autocrine growth. *Science* **245:** 1493-1496.
18. **Rosenthal, S.P.** 1968. Acceleration of primary wound healing by insulin. *Arch. Surg.* **96:** 53–55.
19. **Radulescu, R.T.** 2004. Sequence patterns in the aminoterminus of the tumor suppressor retinoblastoma protein (RB) reminiscent of cyclin E, Bcl-2 and the E7 viral oncoprotein: proposed RB cell survival motifs. *Logical Biol.* **4:** 84-87.
20. **Takahashi, C., and M.E. Ewen.** 2006. Genetic interaction between *Rb* and N-*ras*: differentiation control and metastasis. *Cancer Res.* **66:** 9345-9348.
21. **Radulescu, R.T.** 2006. Insulin-RB heterodimer: potential involvement in the linkage between aging and cancer. *Logical Biol.* **6:** 81-83.
22. **Radulescu, R.T., and C.M. Wendtner.** 1993. Hormone and growth factor subunits: a novel perception of cell growth regulation. *J. Endocrinol.* **139:** 1-7.
23. **Radulescu, R.T.** 2007. Retinoblastoma protein is the likely common effector for distinct anti-aging pathways. *arXiv*:0707.4174v1 [q-bio.SC]
24. **Zhao, L., B. Teter, T. Morihara, G.P. Lim, S.S. Ambegaokar, O.J. Ubeda *et al.*** 2004. Insulin-degrading enzyme as a downstream target of insulin receptor signaling cascade: implications for Alzheimer's disease intervention. *J. Neurosci.* **24:** 11120-11126.
25. **Radulescu, R.T.**, unpublished observation.
26. **Escoubas, P., J.R. De Weille, A. Lecoq, S. Diochot, R. Waldmann, G. Champigny *et al.*** 2000. Isolation of a tarantula toxin specific for a class of proton-gated Na$^+$ channels. *J. Biol. Chem.* **275:** 25116-25121.
27. **Radulescu, R.T.**, unpublished observation.
28. **Xu, J.T., H.Y. Tu, W.J. Xin, X.G. Liu, G.H. Zhang, and C.H. Zhai.** 2007. Activation of phosphatidylinositol 3-kinase and protein kinase B/Akt in dorsal root ganglia and spinal cord contributes to the neuropathic pain induced by spinal nerve ligation in rats. *Exp. Neurol.* **206:** 269-279.
29. **Radulescu, R.T., B.-Y. Zhang, and V. Nüssler.** 1997. RB-like peptides: novel inhibitors of cancer cell growth. *Eur. J. Cell Biol.* **72,** suppl. 44, #40.
30. **Radulescu, R.T., H. Liu, B.-Y. Zhang, W. Wilmanns, and V. Nüssler.** 1997. Retinoblastoma protein-derived peptides: further definition of their antineoplastic properties. *Eur. J. Cancer* **33,** suppl. 5, #83.
31. **Radulescu, R.T.** 1998. Mini-RB versus insulin makes anti-cancer: first positive *in vivo* results with antineoplastic peptides targeting the nucleocrine speedway. *Ann. Oncol.* **9,** suppl. 2, #410.
32. **Radulescu, R.T., and G. Jaques.** 2000. Selective inhibition of human lung cancer cell growth by peptides derived from retinoblastoma protein. *Biochem. Biophys. Res. Commun.* **267:** 71-76.
33. **Radulescu, R.T., C. Cybon, T. Manuilova, and G. Jaques.** 2000. MCR peptides suppress human lung cancer growth *in vitro* and *in vivo*. *Clin. Cancer Res.* **6,** suppl., #487.
34. **Radulescu, R.T., and G. Jaques.** 2003. Potent *in vivo* antineoplastic activity of MCR peptides MCR-4 and MCR-14 against chemotherapy-resistant human small cell lung cancer. *Drugs Exp. Clin. Res.* **29:** 69-74.
35. **Radulescu, R.T., and K. Kehe.** 2003. MCR peptide MCR-4 inhibits the growth of human non-small cell lung cancer through a caspase-independent, p21-dependent pathway. *Logical Biol.* **3:** 45-55.

36. **Radulescu, R.T.**, unpublished observation.
37. **Radulescu, R.T.**, unpublished observation.
38. **Radulescu, R.T.** 1995. From insulin, retinoblastoma protein and the insulin receptor to a new model on growth factor specificity: the nucleocrine pathway. *J. Endocrinol.* **146:** 365-368.
39. **Radulescu, R.T., W. Wilmanns, and V. Nüssler.** 1997. Leucine zipper, nuclear targeting and p130 interaction motifs in the epidermal growth factor precursor: the nucleocrine connection further unveiled. *Eur. J. Cancer* **33,** suppl. 5, #26.
40. **Radulescu, R.T.** 2005. Disconnecting cancer cybernetics through a dual anti-nucleocrine strategy: towards an anticipatory therapy of malignant disease. *Logical Biol.* **5:** 17-29.
41. **Ferrell, J.E.** 1998. How regulated protein translocation can produce switch-like responses. *Trends Biochem. Sci.* **23:** 461-465.
42. **Pollock, R., and M. Gilman.** 1997. Transcriptional activation: is it rocket science? *Proc. Natl. Acad. Sci. USA* **94:** 13388-13389.